\documentclass[epj,nopacs]{svjour}

\usepackage{graphicx}
\usepackage{amsmath}
\usepackage{amssymb}
\usepackage{leftidx}

\newcommand*{\cc}[1]{{#1}^{\ast}}
\newcommand*{\hc}[1]{{#1}^{\dag}}
\newcommand*{\angled}[1]{\left\langle #1 \right\rangle}

\newcommand*{\ximax}{{\xi}_{\mathrm{max}}}

\newcommand*{\ket}[1]{{\left| #1 \right\rangle}}
\newcommand*{\keti}[2]{\ket{#1}_{#2}}
\newcommand*{\bra}[1]{{\left\langle #1 \right|}}
\newcommand*{\brai}[2]{\leftidx{_#2}{\bra{#1}}{}}
\newcommand{\braket}[2]{{\left\langle #1 \middle| #2 \right\rangle}}
\newcommand*{\matrixel}[3]{{\left\langle {#1} \middle| {#2} \middle| {#3} \right\rangle}}
\newcommand*{\abs}[1]{\left| #1 \right|}

\newcommand{\quantity}[3]{#1\times 10^{#2}\;#3}

\newcommand{\per}[1]{#1^{-1}}
\newcommand{\second}{\text{s}}
\newcommand{\Omegai}{\Omega_{0}}
\newcommand{\Omegaf}{\Omega_{\text{f}}}
\newcommand{\rhod}{\rho_{\text{d}}}
\newcommand{\comm}[2]{\left[\, #1 , #2 \,\right]}
\newcommand{\rhos}{\rho_{\text{s}}}
\newcommand{\etal}{et al.}
\newcommand*{\sref}[1]{Section~\ref{#1}}
\newcommand*{\Fref}[1]{Figure~\ref{#1}}
\newcommand{\rme}{\mathrm{e}}
\newcommand{\rmi}{\mathrm{i}}

\newcommand{\eref}[1]{\ref{#1}}

\newcommand*\f[1]{\fin#1\relax}
\def\fin#1,#2\relax{f_{#1,#2}}
\newcommand*\fcc[1]{\fccin#1\relax}
\def\fccin#1,#2\relax{\cc{f}_{#1,#2}}

\newcommand*{\calE}[2]{\mathcal{E}^{(#1)}_{#2}}

\begin{document}

\title{The effect of disorder on polaritons in a coupled array of cavities}
\author{Abuenameh Aiyejina \and Roger Andrews}
%
%
\institute{The Department of Physics, The University of the West Indies, St. Augustine, Trinidad and Tobago}
\date{}
%
\abstract{
The effect of disorder in the intensity of the driving laser on a coupled array of cavities described by a Bose-Hubbard Hamiltonian for dark-state polaritons is investigated. A canonically-transformed Gutzwiller wave function is used to investigate the phase diagram and dynamics of a one-dimensional system with uniformly distributed disorder in the Rabi frequency. In the phase diagram, we find the emergence of a Bose glass phase that increases in extent as the strength of the disorder increases. We study the dynamics of the system when subject to a ramp in the Rabi frequency which, starting from the superfluid phase, is decreased linearly and then increased to its initial value. We investigate the dependence of the density of excitations, the relaxation of the superfluid order parameter and the excess energy pumped into the system on the inverse ramp rate, $\tau$. We find that, in the absence of disorder, the defect density oscillates with a constant envelope, while the relaxation of the order parameter and excess energy oscillate with $\tau^{-1.5}$ and $\tau^{-2}$ envelopes respectively. In the presence of disorder in the Rabi frequency, the defect density oscillates with a decaying envelope, the relaxation of the order parameter no longer decreases as $\tau$ increases while the residual energy decreases as $\tau$ increases. The rate at which the envelope of the defect density decays increases with increasing disorder strength, while the excess energy falls off more slowly with increasing disorder strength.
}

\maketitle

\section{\label{sec:intro}Introduction}

The Bose-Hubbard model has received significant theoretical and experimental study as a model of quantum many-body phenomena since the seminal paper by Fisher \etal~\cite{Fisher:1989il}. In that paper, it was shown that at zero temperature and in the absence of disorder, a system described by the Bose-Hubbard model exhibits two phases -- the Mott insulator phase and the superfluid phase. Fisher \etal also showed that the addition of disorder to the Bose-Hubbard model gives rise to a third phase -- the Bose glass phase. This phase is insulating due to the localizing effect of disorder, but is compressible and gapless like the superfluid phase.

The disordered Bose-Hubbard model has been investigated with various approaches, including field-theoretic techniques~\cite{Svistunov:1996oz,Pazmandi:1998xr,Hastings:2001pi}, quantum Monte Carlo simulations~\cite{Kisker:1997mi,Lee:2004fe,Hitchcock:2006ec,Roscilde:2008la}, DMRG~\cite{Goldsborough:2015kx} and mean-field approximations~\cite{Krutitsky:2006bs,Bissbort:2009ij,Bissbort:2010fk}. The site-dependent Gutzwiller approximation or the equivalent site-decoupling mean field approximation and the time-dependent Gutzwiller approximation have been used to study the phase diagram~\cite{Buonsante:2007fv,Buonsante:2007th,Buonsante:2007bd,Buonsante:2009ss} and dynamics~\cite{Jaksch:1998xw,Damski:2003ye,Zakrzewski:2005uf} of the Bose-Hubbard model. In an extension to the Gutzwiller technique, Lin \etal~\cite{Lin:2012uq} investigate the phase diagram and dynamics of the two-dimensional Bose-Hubbard model with an on-site disorder potential using a variational wave function approach with a canonical transformation. We use a similar transformation in this paper for a one-dimensional system. This transformation incorporates quantum fluctuations over the mean field theory of the Gutzwiller technique and builds in non-local correlations.

The Bose-Hubbard model has been experimentally realized using arrays of Josephson junctions~\cite{Fazio:2001fk} and cold atoms in optical lattices~\cite{Bloch:2008uq}. A great deal of work has focused on atoms in optical lattices since this system provides a defect-free lattice and control over the ratio of the on-site repulsion to the hopping strength via the laser intensity~\cite{Jaksch:1998xw}. Different approaches have been used to introduce disorder into this system including the addition of an incommensurate lattice~\cite{Roth:2003zv,Damski:2003ye,Fallani:2007jl} and the use of a laser speckle potential~\cite{Lye:2005il,Schulte:2005gb,Clement:2005zt}.

In this paper, we will examine the effect of disorder on a system proposed by Hartmann \etal~\cite{Hartmann:2006sj,Hartmann:2008sw}. This system consists of an array of coupled optical cavities, each containing a large number of four-level atoms that are driven by an external laser with uniform intensity across the cavities. A brief review of the properties of coupled quantum electrodynamics cavities was done by Tomadin and Fazio~\cite{Tomadin:2010nr}. Hartmann \etal found that under certain conditions this system can be described by a Bose-Hubbard model for combined atom-photon excitations called polaritons and that it exhibits a Mott-insulator-to-superfluid transition. Rossini and Fazio~\cite{Rossini:2007ud} examined the phase diagram of this system in one dimension using the density matrix renormalization group (DMRG) technique. They found that for the phase diagram, the Bose-Hubbard model is a good approximation for this system as long as the number of atoms in each cavity is sufficiently large. Even values $\sim 10$ for the number of atoms were sufficient. They also looked at the effects of random fluctuations in the number of atoms per cavity. In this paper we consider the case of non-uniform intensity of the external laser and investigate the resulting phase diagrams. We also examine the effect of an external laser with a time-dependent, non-uniform intensity on the dynamics of the system.

In \sref{sec:bhm}, we introduce the system proposed by Hartmann \etal and we show its extension to the case of disorder in \sref{sec:disorder}. In \sref{sec:can} we give the canonical transformation that was employed and in \sref{sec:phase} we show how this transformation can be used to obtain a qualitative phase diagram of a one-dimensional system. Following that, the dynamics problem that was investigated is outlined in \sref{sec:dynamics}. In \sref{sec:phaseresults}, we obtain the phase diagram of a one-dimensional system of 25 cavities in the presence of disorder in the laser intensity and in \sref{sec:dynamicsresults} we look at the dynamics of this system subject to a ramp in the Rabi frequency of the driving by the laser. For the dynamics of our one-dimensional polariton system, we investigate the defect density, the relaxation of the superfluid order parameter and the excess energy pumped into the system as a result of the ramp.

\section{\label{sec:theory}Theory}

\subsection{\label{sec:bhm}The Bose-Hubbard model}

The Bose-Hubbard Hamiltonian for bosons on a lattice is given by
\begin{equation}
H = \frac{1}{2}U\sum_{i}{n_i\left(n_i - 1\right)} - J\sum_{\angled{ij}}{\left(\hc{a_i}a_j + \hc{a_j}a_i\right)},\label{eq:BH}
\end{equation}
where $\hc{a_i}$ ($a_i$) is the creation (annihilation) operator for bosons at lattice site $i$, $n_i = \hc{a_i}a_i$ is the number operator for bosons at lattice site $i$ and $\angled{ij}$ indicates a sum over pairs of adjacent sites $i$ and $j$. Here $U$ is the strength of the on-site repulsion between bosons at a given site and $J$ is the strength of the hopping of bosons between adjacent sites.

As described by Hartmann \etal~\cite{Hartmann:2006sj}, the polariton system consists of an array of optical cavities each containing $N$ ($\gg$ 1) four-level atoms. The cavities are sufficiently close together so that there is an overlap of the evanescent electromagnetic fields of adjacent cavities. Each cavity has a resonance frequency $\omega$, and the overlap integral for the electromagnetic modes of adjacent cavities is given by $\alpha$. The strengths of the couplings between the electromagnetic mode in a cavity and the transitions between atomic levels 1 and 3 and levels 2 and 4 are given by dipole coupling parameters $g_{13}$ and $g_{24}$ respectively. An external laser drives the transition between levels 2 and 3 with Rabi frequency $\Omega$. The detunings from levels 2, 3 and 4 are given by $\varepsilon$, $\delta$ and $\Delta$ respectively. When the couplings, Rabi frequency and detunings satisfy certain conditions, and cavity and atomic decay are neglected, the system can be described by a Bose-Hubbard model for dark-state polaritons. In this case the on-site repulsion and hopping strength are given by
\begin{equation}
U = -\frac{2g_{24}^2}{\Delta}\frac{Ng_{13}^2\Omega^2}{\left(Ng_{13}^2 + \Omega^2\right)^2}
\quad \text{and} \quad
J = \frac{2\omega\alpha\Omega^2}{Ng_{13}^2 + \Omega^2}.
\end{equation}

\subsection{\label{sec:disorder}Disordered Polariton System}

In the presence of disorder in the on-site repulsion and hopping strength, the Hamiltonian of the system becomes
\begin{equation}
H = \frac{1}{2}\sum_{i}{U_i n_i\left(n_i - 1\right)} - \sum_{\angled{ij}}{J_{ij}\left(\hc{a_i}a_j + \hc{a_j}a_i\right)},\label{eq:BHdis}
\end{equation}
where $U_i$ is the strength of the on-site repulsion at site $i$ and $J_{ij}$ is the strength of the hopping between sites $i$ and $j$. For the case of disorder in the Rabi frequency, the on-site repulsion and hopping strength are given by
\begin{subequations}
\begin{equation}
U_i = -\frac{2g_{24}^2}{\Delta}\frac{Ng_{13}^2\Omega_i^2}{\left(Ng_{13}^2 + \Omega_i^2\right)^2}
\end{equation}
and
\begin{equation}
J_{ij} = \frac{2\omega\alpha\Omega_i\Omega_j}{\sqrt{Ng_{13}^2 + \Omega_i^2\vphantom{\Omega_j^2}}\sqrt{Ng_{13}^2 + \Omega_j^2}},
\end{equation}
\end{subequations}
where $\Omega_i$ denotes the value of the Rabi frequency at cavity $i$. In this paper, we consider the case of uniform disorder in the Rabi frequency. The Rabi frequency at cavity $i$ is given by $\Omega_i = \Omega(1 + \xi_i)$, where $\Omega$ is the mean Rabi frequency and $\xi_i$ is an uncorrelated random variable uniformly distributed in the interval $\left[-\ximax, \ximax\right]$.

\subsection{\label{sec:can}The Canonical Transformation}

In order to study the phase diagram and dynamics of the disordered Bose-Hubbard model, we use a canonical transformation in the vein of Lin \etal~\cite{Lin:2012uq}. The canonical transformation is used to partially remove the high-energy terms from the Hamiltonian. The remaining high-energy terms are small compared to the low-energy terms and will be shown theoretically to be the case at the end of this section. For the phase diagram, we use as a variational wave function that consists of a canonically-transformed Gutzwiller wave function of the form
\begin{equation}
\ket{\psi} = \rme^{-\rmi\mathcal{S}}\ket{\psi_0}, \quad \ket{\psi_0} = \bigotimes_i{\ket{\psi_0^i}}, \quad \ket{\psi_0^i} = \sum_{n}{\f{i,n}\keti{n}{i}}.
\end{equation}
Here $\keti{n}{i}$ is the number state corresponding to $n$ particles on site $i$, $\f{i,n}$ is the Gutzwiller coefficient corresponding to the state $\keti{n}{i}$ and $\rme^{-\rmi\mathcal{S}}$ is a canonical transformation that builds in nonlocal correlations in the proximity of a Mott insulator. The transformation will be chosen such that it removes most of the high-energy terms in the Hamiltonian and further details are given in Appendix A.

An operator $A$ is canonically transformed to $A^\ast$ by the formula $A^\ast = \rme^{\rmi\mathcal{S}}A\rme^{-\rmi\mathcal{S}}$. The Gutzwiller wave function can be written in terms of the variational state by inverting the transformation to give $\ket{\psi_0} = \rme^{\rmi\mathcal{S}}\ket{\psi}$. The expectation value, $\matrixel{\psi_0}{A^\ast}{\psi_0}$, of the transformed operator, $A^\ast$, in the transformed variational state, $\ket{\psi_0}$,  is therefore equal to the expectation value of the original operator in the variational state, $\matrixel{\psi}{A}{\psi}$.

In order to perform the canonical transformation, the Hamiltonian \eref{eq:BHdis} is split into a local Hamiltonian and the hopping term:
\begin{align}
H &= H_0 + \sum_{\angled{ij}}{T_{ij}}, \quad H_0 = \sum_{i}{\frac{1}{2}U_i n_i\left(n_i - 1\right) - \mu n_i}, \notag\\
T_{ij} &= -J_{ij}\hc{a_i}a_j,
\end{align}
where the chemical potential $\mu$ has been added. We can decompose the hopping term as follows:
\begin{equation}
T_{ij} = \sum_{nm}{T_{ij}^{nm}}, T_{ij}^{nm} = -J_{ij}g_{nm}\keti{n+1}{i}\keti{m-1}{j}\brai{n}{i}\brai{m}{j},
\end{equation}
where $g_{nm} = \sqrt{m(n+1)}$. The hopping term $T_{ij}^{nm}$ connects states differing in local energy by $\varepsilon_{ij}^{nm} = n U_i - (m-1)U_j$, where $n$ is the initial number of particles at site $i$ and $m$ is the initial number of particles at site $j$. If we write the on-site repulsion as $U_i = U_0 + \delta U_i$, then we have $\varepsilon_{ij}^{nm} = \varepsilon_0^{nm} + \delta\varepsilon_{ij}^{nm}$, where $\varepsilon_0^{nm} = (n - m + 1)U_0$ and $\delta\varepsilon_{ij}^{nm} = n\delta U_i - (m-1)\delta U_j$.

In order to determine the canonical transformation, we assume that we can expand the canonical transformation operator $\rmi\mathcal{S}$ in powers of $J/U$ as $\rmi\mathcal{S} = \rmi\mathcal{S}^1 + \rmi\mathcal{S}^2 + \ldots$, where $\rmi\mathcal{S}^m \sim (J / U)^m$. As shown in Appendix A, the first two terms in this expansion are given by
\begin{equation}
\rmi\mathcal{S}^1 = \sum_{\angled{ij}}\sum_{n\ne m-1}\frac{T_{ij}^{nm}}{\varepsilon_0^{nm}}
\end{equation}
and
\begin{align}
\rmi\mathcal{S}^2 &= \sum_{\angled{ij}\angled{kl}}{\sum_{\substack{n\ne m-1\\p}}{\frac{1}{\left(\varepsilon_0^{nm}\right)^2}\comm{T_{ij}^{nm}}{T_{kl}^{p,p+1}}}}\notag\\
&\qquad + \frac{1}{2}\sum_{\angled{ij}\angled{kl}}{\sum_{\substack{n\ne m-1\\p\ne q-1\\n-m\ne p-q}}{\frac{\comm{T_{ij}^{nm}}{T_{kl}^{q-1,p+1}}}{\varepsilon_0^{nm}\left(\varepsilon_0^{nm} + \varepsilon_0^{q-1,p+1}\right)}}}.
\end{align}
We were unable to remove all high-energy terms because it was necessary to leave out the disorder in $U$ in the canonical transformation in order for the matrix exponential $\exp(\rmi\mathcal{S})$ to converge sufficiently quickly.

These expressions can be used to canonically transform the Hamiltonian using the nested commutator expansion. This gives
\begin{align}
H^\ast &= H_0 + \sum_{\angled{ij}}\sum_{n}T_{ij}^{n,n+1}\notag\\
& + \frac{1}{2}\sum_{\angled{ij}\angled{kl}}\sum_{\substack{n\ne m-1\\p\ne q-1\\n-m=p-q}}\frac{1}{\varepsilon_0^{nm}}\comm{T_{ij}^{nm}}{T_{kl}^{q-1,p+1}}\notag\\
& - \sum_{\angled{ij}}\sum_{n\ne m-1}\frac{\delta\varepsilon_{ij}^{nm}}{\varepsilon_0^{nm}}T_{ij}^{nm}\notag\\
& - \frac{1}{2}\sum_{\angled{ij}\angled{kl}}\sum_{\substack{n\ne m-1\\p\ne q-1}}\frac{\delta\varepsilon_{kl}^{q-1,p+1}}{\varepsilon_0^{nm}\varepsilon_0^{q-1,p+1}}\comm{T_{ij}^{nm}}{T_{kl}^{q-1,p+1}}\notag\\
& - \frac{1}{2}\sum_{\angled{ij}\angled{kl}}\sum_{\substack{n\ne m-1\\p}}\frac{\delta\varepsilon_{ij}^{nm}+\delta\varepsilon_{kl}^{p,p+1}}{\left(\varepsilon_0^{nm}\right)^2}\comm{T_{ij}^{nm}}{T_{kl}^{p,p+1}}\notag\\
& - \frac{1}{2}\sum_{\angled{ij}\angled{kl}}\sum_{\substack{n\ne m-1\\n-m\ne p-q}}\frac{\delta\varepsilon_{ij}^{nm}+\delta\varepsilon_{kl}^{q-1,p+1}}{\varepsilon_0^{nm}\left(\varepsilon_0^{nm} + \varepsilon_0^{q-1,p+1}\right)}\notag\\
&\times\comm{T_{ij}^{nm}}{T_{kl}^{q-1,p+1}}.
\end{align}
The first three terms are the low-energy terms. The remaining terms are the residual high-energy terms, which are less significant due to the terms involving $\delta\varepsilon$.

\section{\label{sec:methods}Methods}

\subsection{\label{sec:phase}Phase diagram}

In order to find the ground state of the Bose-Hubbard Hamiltonian, the expectation value, $\mathcal{E}$, of the effective Hamiltonian $H^\ast$ in the Gutzwiller state $\ket{\psi_0}$ was calculated. The expression for $\mathcal{E}$ is given in \ref{app:gsenergy}. The ground state was determined by numerically minimizing $\mathcal{E}$ with respect to the variables $\f{i,n}$. In order to determine the phase of the system, the superfluid stiffness is needed. To calculate this, Peierls phase factors were added to the hopping parameter, $J_{ij} \rightarrow J_{ij}\rme^{\rmi\theta\left(x_i - x_j\right)}$, where $x_i$ and $x_j$ are lattice positions and $\theta$ is a small phase~\cite{Lin:2012uq}. This phase corresponds to applying a small phase gradient across the lattice to induce a superfluid flow. The corresponding ground state energy of the new Hamiltonian, $\mathcal{E}_{\theta}$, was found. The superfluid stiffness was then calculated as
\begin{equation}
\rhos = \frac{1}{N_{\mathcal{C}}}\sum_{\mathcal{C}}\frac{\mathcal{E}_{\theta} - \mathcal{E}}{\theta^2},
\end{equation}
where $\mathcal{C}$ denotes disorder realizations and $N_{\mathcal{C}}$ is the number of disorder realizations. In addition to the superfluid stiffness, the coefficients $\f{i,n_0}$ were used to determine the phase, where $n_0$ is the occupation number of each site in the Mott lobe under consideration. In this paper, we are looking at the first Mott lobe, so we take $n_0 = 1$. Any point in the phase diagram with non-zero superfluid stiffness was identified as being in the superfluid phase. Among the points with zero superfluid stiffness, those for which $\f{i,n_0} = 1$ for all lattice sites in all disorder configurations were identified as the Mott insulator phase, and those for which $\f{i,n_0} < 1$ for at least one site in some disorder configuration were identified as the Bose glass phase.~\cite{Lin:2012uq}

\subsection{\label{sec:dynamics}Dynamics}

In order to study the dynamics of the system, we use a variational wave function of the form
\begin{align}
\ket{\psi(t)} &= \rme^{-\rmi\mathcal{S}[\Omega(t)]}\ket{\psi_0(t)},\notag\\
\ket{\psi_0(t)} &= \bigotimes_i{\ket{\psi_0^i(t)}}, \quad \ket{\psi_0^i(t)} = \sum_{n}{\f{i,n}(t)\keti{n}{i}},
\end{align}
where the canonical transformation is evaluated with the instantaneous value of the Rabi frequency, $\Omega(t)$. From this the Schr\"{o}dinger equation becomes
\begin{equation}
\rmi\dot{\ket{\psi_0}} = \left(H^{\ast} - \dot{\mathcal{S}}^{\ast}\right)\ket{\psi_0},
\end{equation}
where $\dot{\mathcal{S}^{\ast}} = \rme^{\rmi\mathcal{S}}\dot{\mathcal{S}}\rme^{-\rmi\mathcal{S}}$. As in Lin \etal~\cite{Lin:2012uq}, we keep only the first-order term, $\dot{\mathcal{S}^1}$ and, since $\rmi\dot{\mathcal{S}^1} \propto \rmi\mathcal{S}^1$, we are also able to make the simplification $\rmi\dot{\mathcal{S}}^{\ast} = \rmi\dot{\mathcal{S}}$. The time-dependent variational principle was used to obtain a system of differential equations that were then solved using a fourth-order Runge-Kutta method.

We investigated the dynamics of the system subjected to a ramp in the Rabi frequency. The system was started in its ground state for an initial value of the mean Rabi frequency, $\Omegai$, which puts it in the superfluid phase. We only consider disorder realizations where the ground state wavefunction is concentrated in the $f_{i,1}$ and $f_{i,2}$ coefficients as we are looking at the first Mott lobe. The Rabi frequency was then decreased linearly to a value of $\Omegaf$ at a rate $\tau^{-1}$, where $\tau$ is the inverse ramp rate. Next, the Rabi frequency was increased linearly back to $\Omegai$ at the same rate. The evolution of the Rabi frequency is thus given by
\begin{equation}
\Omega(t) =
\begin{cases}
\Omegai  + \left(\Omegaf - \Omegai\right)\frac{t}{\tau},& t \leqslant \tau\\
\Omegaf  + \left(\Omegai - \Omegaf\right)\frac{t-\tau}{\tau},& t \geqslant \tau.
\end{cases}
\end{equation}

Motivated by Lin \etal, we looked at the defect density, $\rhod$, the superfluid order parameter, $\Phi$, and the residual energy, $Q$. The defect density is the number density of excitations created in the system due to the non-equilibrium ramp and is given by
\begin{equation}
\rhod(\tau) = \frac{1}{L}\sum_i{p_i}, \quad p_i = 1 - \abs{\braket{\psi_0^i(2\tau)}{\psi_0^i(0)}}^2.
\end{equation}
The superfluid order parameter is given by
\begin{equation}
\Phi(t) = \frac{1}{L}\sum_i{\matrixel{\psi(t)}{a_i}{\psi(t)}}.
\end{equation}
We examined the relaxation of the superfluid order parameter to its initial value by looking at the quantity $\abs{\abs{r} - 1}$ where $r = \Phi(2\tau)/\Phi(0)$ is the normalized final value of the order parameter. We also examined the excess energy pumped into the system in the process of ramping down the Rabi frequency and ramping it back up to the initial value. The excess energy in the final state is given by
\begin{equation}
Q = \matrixel{\psi(2\tau)}{H}{\psi(2\tau)} - \matrixel{\psi(0)}{H}{\psi(0)}.
\end{equation}

\section{\label{sec:results}Results and Discussion}

\subsection{\label{sec:phaseresults}Phase Diagram}

Hartmann \etal~\cite{Hartmann:2006sj} gave a set of parameters for which the Bose-Hubbard model is a good approximation for the polariton system. These parameters were $g_{13} = g_{24} = \quantity{2.5}{9}{\per\second}$, $\varepsilon = 0$, $\delta = \quantity{1.0}{12}{\per\second}$, $\Delta = \quantity{-2.0}{10}{\per\second}$, $N = 1000$, and $2\omega\alpha = \quantity{1.1}{7}{\per\second}$ and they are used throughout this paper. In addition, the sum in the Gutzwiller wave function was cut-off at $n = 7$, since the Gutzwiller coefficients for larger $n$ were negligible. For the calculations, the Hamiltonian was scaled by the value of the on-site repulsion corresponding to the mean Rabi frequency $\Omega$. We then set $U_0 = 1$.

\begin{figure*}
\begin{center}
\includegraphics[width=\textwidth]{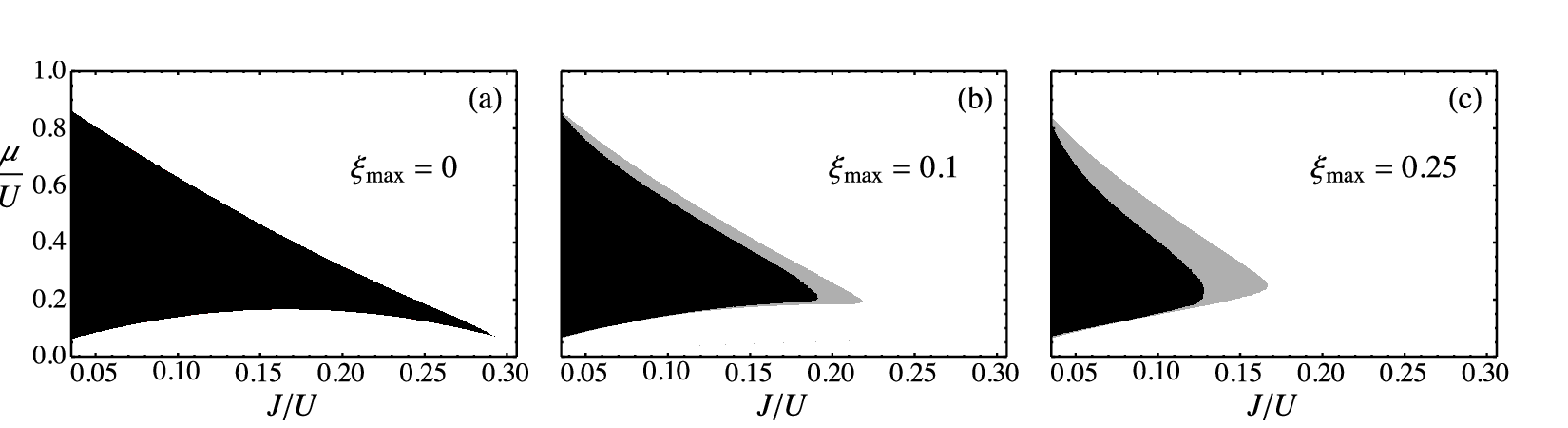}
\end{center}
\caption{Phase diagrams for a system of 25 cavities with (a) no disorder, (b) uniform disorder in the Rabi frequency with amplitude $\ximax = 0.1$ and (c) uniform disorder with amplitude $\ximax = 0.25$, where $\mu$ is the chemical potential and $J/U$ is the ratio of the hopping strength to the on-site repulsion corresponding to the Rabi frequencies used. The black region is the Mott insulator phase, the gray region is the Bose glass phase and the white region is the superfluid phase.\label{fig:Phase}}
\end{figure*}

The phase diagram was found for a one-dimensional system of 25 cavities with periodic boundary conditions. Initially, the phase diagram of the system was determined in the absence of disorder. Figure \ref{fig:Phase} (a) shows the phase diagram obtained for the region $\quantity{2}{10}{\per\second} \le \Omega \le \quantity{3.2}{11}{\per\second}$ and $0 \le \mu \le 1$, which partially covers the first Mott lobe. The minimum value of $J/U$, corresponding to $\Omega = \quantity{2}{10}{\per\second}$, is $\sim 0.0187$. The black region is the Mott insulator phase and the white region is the superfluid phase. The phase diagram is qualitatively similar to that obtained using DMRG~\cite{Ejima:2011fk} with a similar value for the Mott tip of $\left(J/U\right)_\mathrm{c} \approx 0.293$. As a result of the non-zero minimum value of $J/U$, the Mott lobe does not extend all the way to $\mu/U = 0$ and $\mu/U = 1$.

The phase diagram was also obtained for the case of uniform disorder in the Rabi frequency. Figures \ref{fig:Phase} (b) and (c) show the phase diagrams obtained for two different disorder amplitudes, $\ximax = 0.1$ and $\ximax = 0.25$. Each phase diagram was obtained by averaging over 100 realizations of uniform disorder in $\Omega$. The black region is the Mott insulator phase, the gray region is the Bose glass phase and the white region is the superfluid phase.

The addition of disorder causes the appearance of the Bose glass phase around the Mott insulator phase. As expected, the extent of this phase was greater for the larger disorder strength. Due to the non-zero minimum value of $J/U$, the superfluid phase persists to the y-axis for $\ximax = 0.1$. The tip of the Mott lobe moves to smaller values of $J/U$ for increased disorder. The canonical transformation used is more accurate for smaller disorder amplitudes, so for $\ximax = 0.25$ the numerical accuracy of the results would have been negatively affected.

The simulations were also run with the disorder in $U$ artificially suppressed and then with the disorder in $J$ suppressed. When the disorder in $U$ was suppressed, the phase diagrams were the same as they were without disorder. Any Bose glass phase was too small in extent to appear on the phase diagram. On the other hand, when the disorder in $J$ was suppressed, the phase diagrams were the same as they were with disorder in both $J$ and $U$. This is due to the fact that the disorder in $J$ that is induced by the disorder in $\Omega$ is quite small, with a maximum deviation of $\sim 2\%$ for $\ximax = 0.1$ and $\sim 7\%$ for $\ximax = 0.25$. On the other hand, the disorder in $U$ has a maximum deviation of $\sim 20\%$ for $\ximax = 0.1$ and $\sim 55\%$ for $\ximax = 0.25$. Thus the phase diagrams obtained are determined almost solely by the disorder in the on-site repulsion.

Our results are qualitatively similar to those of Gimperlein \etal~\cite{Gimperlein:2005fk} as we find that, with disorder, the Mott lobe shrinks in both the $\mu$ and $J$ directions and the lower boundary of the Mott lobe doesn't vary significantly with the disorder amplitude up to the critical hopping. However, under our approximation, we were unable to obtain the same behaviour for the tip of the lobe, especially for the larger disorder amplitude. Gimperlein \etal used a strong coupling expansion and quantum Monte Carlo simulations to investigate the effect of disorder in the on-site interactions on a Bose-Hubbard model for ultracold atoms in an optical lattice.

\subsection{\label{sec:dynamicsresults}Dynamics}

For the dynamics calculations in this paper, we used $\Omegai = \quantity{3}{11}{\per\second}$, $\Omegaf = \quantity{1}{11}{\per\second}$, and $\mu = 0.5U$. The calculations were done for inverse ramp rates, $\tau$, ranging from $\quantity{1}{-7}{\second}$ to $\quantity{2}{-6}{\second}$.  These values should take us into the large $\tau$ limit as $\tau U \gg 1$ for the longest ramp times used. We considered the non-disordered case and the cases of uniform disorder in the Rabi frequency with amplitudes $\ximax = 0.1$ and $\ximax = 0.25$. In the disordered case, the calculations were done for 200 realizations of uniform disorder in $\Omega$. We ran the results from 100 realizations up to 200 and found good convergence for 200 realizations. For example, the percentage change going from 199 to 200 realizations for the defect density was less than 1\%. The other parameters show even smaller changes. The defect density, $\rhod$, residual energy, $Q$, and $\abs{r}$ were obtained by averaging over the realizations. The quantity $\abs{\abs{r}-1}$ was calculated from the disorder-averaged value of $\abs{r}$. 

\begin{figure*}
\begin{center}
\includegraphics[width=\textwidth]{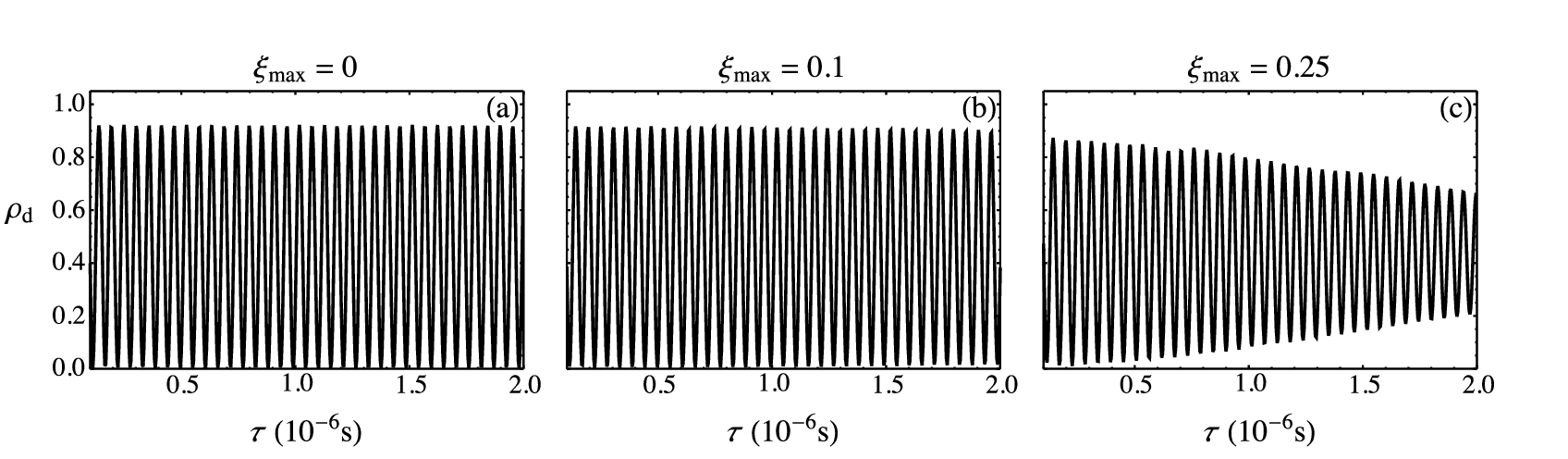}
\end{center}
\caption{The defect density as a function of the inverse ramp rate, $\tau$, in the case of (a) no disorder, (b) uniform disorder in the Rabi frequency with amplitude $\ximax = 0.1$ and (c) disorder with amplitude $\ximax = 0.25$.\label{fig:rhoplots}}
\end{figure*}

\begin{figure*}
\begin{center}
\includegraphics[width=\textwidth]{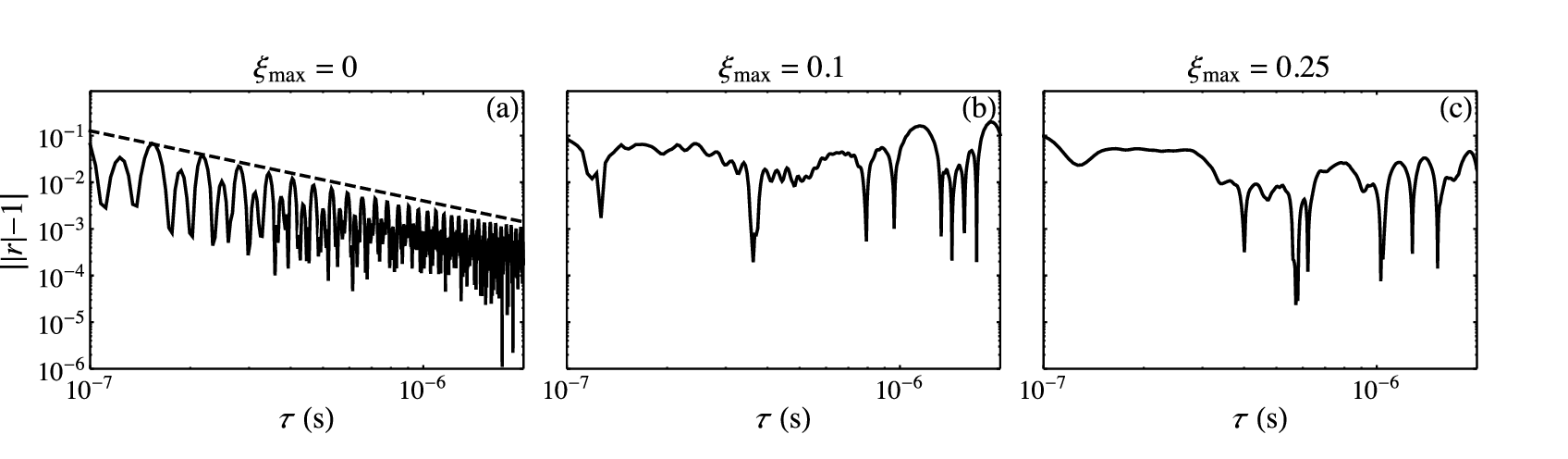}
\end{center}
\caption{The relaxation of the superfluid order parameter as a function of the inverse ramp rate, $\tau$, in the case of (a) no disorder, (b) uniform disorder in the Rabi frequency with amplitude $\ximax = 0.1$ and (c) disorder with amplitude $\ximax = 0.25$. A dashed line has been added to the first figure to show the $\tau^{-1.5}$ dependence.\label{fig:rplots}}
\end{figure*}

\begin{figure*}
\begin{center}
\includegraphics[width=\textwidth]{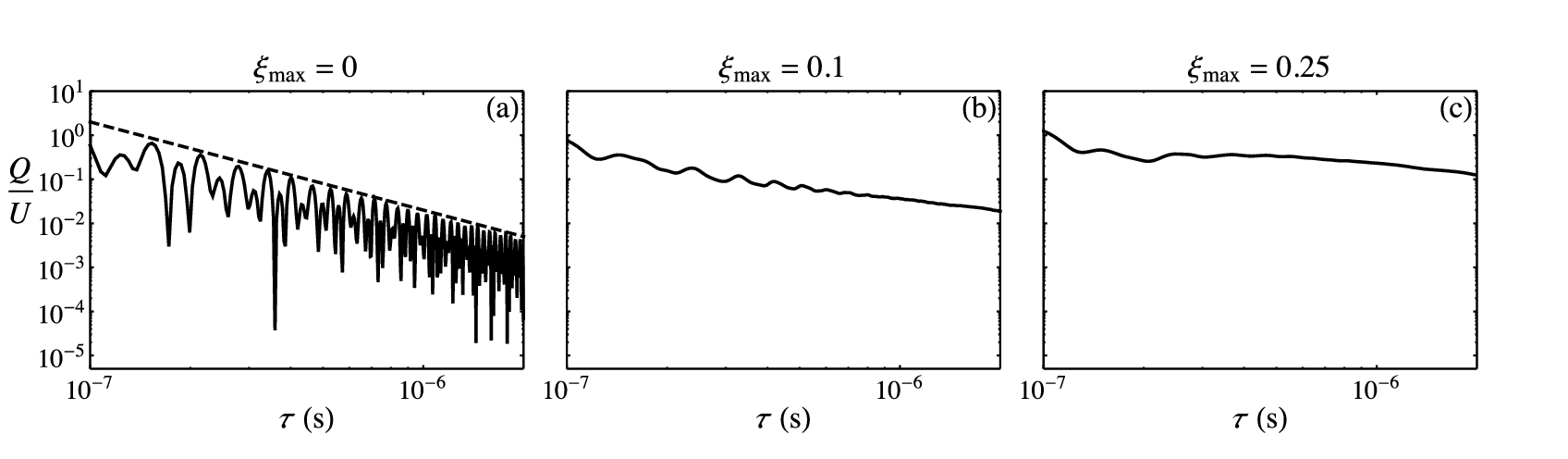}
\end{center}
\caption{The excess energy as a function of the inverse ramp rate, $\tau$, in the case of (a) no disorder, (b) uniform disorder in the Rabi frequency with amplitude $\ximax = 0.1$ and (c) disorder with amplitude $\ximax = 0.25$. A dashed line has been added to the first figure to show the $\tau^{-2}$ dependence.\label{fig:Qplots}}
\end{figure*}

\Fref{fig:rhoplots} shows the results for the defect density in the three cases considered. \Fref{fig:rhoplots} (a) shows the results in the absence of disorder. The defect density was found to oscillate with a constant envelope, with the system returning to the initial state ($\rhod = 0$) for certain values of $\tau$. Figures \ref{fig:rhoplots} (b) and (c) show the results for disorder. In both cases the defect density shows oscillations similar to the clean case. For disorder strength $\ximax = 0.1$, the oscillations have almost constant amplitude, with just a very slight decrease over the range considered. For $\ximax = 0.25$, there is a clear reduction in the amplitude of the oscillations and the amplitude falls off with increasing $\tau$.

\Fref{fig:rplots} shows the results for the relaxation of the superfluid order parameter. \Fref{fig:rplots} (a) shows the results for the non-disordered case. The quantity $\abs{\abs{r}-1}$ exhibits oscillations with a roughly $\tau^{-1.5}$ envelope. A dashed line has been added to the figure to illustrate the $\tau^{-1.5}$ dependence. When disorder was added, the results in Figures \ref{fig:rplots} (b) and (c) were obtained. The values of $\abs{\abs{r}-1}$ show irregular oscillations in these cases, and there is no overall decrease in the values as $\tau$ increases. The disorder therefore destroys the scaling with $\tau$. 

\Fref{fig:Qplots} show the results for the excess energy. In the case of no disorder, \Fref{fig:Qplots} (a) shows that the excess energy has oscillations with a $\tau^{-2}$ envelope. Again, a dashed line has been added to the figure to illustrate the $\tau^{-2}$ dependence. In the presence of disorder, Figures \ref{fig:Qplots} (b) and (c) show that the excess energy decreases with increasing $\tau$, but shows much less oscillation than the case of no disorder. The excess energy falls off more slowly for the larger disorder strength. 

If the evolution of this system is adiabatic, we would expect that $\abs{\abs{r}-1}$ and $Q/U$ would approach 0 in the large $\tau$ limit as the initial and final energies and superfluid order parameters would be the same. In the absence of disorder, at the largest value of $\tau$ considered, $Q/U \sim 10^{-2}$ while in the presence of disorder $Q/U \sim 10^{-1}$. Also, in the absence of disorder, we find that $\abs{\abs{r}-1} \sim 10^{-3}$ in the large $\tau$ limit. However, in the presence of disorder, $\abs{\abs{r}-1} \sim 10^{-1}$. While it is possible that in the absence of disorder the evolution is adiabatic in the large $\tau$ limit, the addition of disorder appears to reduce any adiabaticity.

In the paper by Lin \etal~\cite{Lin:2012uq}, they found that for a two dimensional system without disorder, the relaxation of the superfluid order parameter has a $\tau^{-1}$ envelope. They link this to the lowest order dependence of the energy on the superfluid order parameter in a Gross-Pitaevskii description. In contrast, for a one-dimensional system we find a $\tau^{-1.5}$ envelope. However, the large $\tau$ behaviour of the excess energy remains the same. This result is relevant to theory and experiment since the large $\tau$ behaviour of the relaxation of the superfluid order parameter differs between the one-dimensional and two-dimensional systems.

\section{\label{sec:conclusion}Conclusion}

In this paper we used a canonical transformation technique to investigate the phase diagram as well as the dynamics of a disordered Bose-Hubbard model realized by polaritons in a coupled array of cavities. The disorder was introduced into the Rabi frequency at different sites using uniformly distributed disorder. We used a variational wave function given by a canonically-transformed Gutzwiller wavefunction. The phase diagrams showed the expected Mott insulator, Bose glass and superfluid phases, with the extent of the Bose glass phase increasing as the strength of the disorder increased. In the absence of disorder, the phase diagram was qualitatively similar to the phase diagram obtained with DMRG.

The dynamics were investigated for the case where the system was started from a ground state in the superfluid phase and the Rabi frequency was linearly ramped down and then back up at a rate $\tau^{-1}$. We looked at the relaxation of the superfluid order parameter as well as the density of excitations created and residual energy pumped into the system by the ramp. In the absence of disorder, all three quantities showed oscillations as a function of the inverse ramp rate $\tau$. The envelope of the defect density was constant while the relaxation of the order parameter and the residual energy had $\tau^{-1.5}$ and $\tau^{-2}$ envelopes respectively.

In the presence of disorder, the defect density still showed oscillations, but the amplitude of the oscillations decreased for increasing $\tau$. The rate at which the oscillations were damped increased as the strength of the disorder increased. The relaxation of the superfluid parameter exhibited some irregular oscillations and a lack of overall decrease in the large $\tau$ limit. The excess energy showed fewer oscillations which had smaller amplitudes than in the absence of disorder, but still decreased as $\tau$ increased. The fall-off of the excess energy with respect to $\tau$ became slower as the strength of disorder increased.

\section*{Author contribution statement}

Abuenameh Aiyejina performed the simulations. Both authors wrote the manuscript together.

\appendix

\section{\label{app:can}The Canonical Transformation}

We wish to investigate the low energy behaviour of the disordered Bose-Hubbard Hamiltonian. To this end we develop a canonical transformation that removes the terms that contribute to high-energy transitions. Hopping terms of the form $T_{ij}^{nm}$, when $n\ne m-1$, connect states that differ in energy by $\sim U$ or more. So we will create a transformation that removes these terms.

The canonically-transformed Hamiltonian is given by $H^{\ast} = \rme^{\rmi\mathcal{S}}H\rme^{-\rmi\mathcal{S}}$. This can be expanded using nested commutators to give
\begin{align}
H^{\ast} &= H + \comm{\rmi\mathcal{S}}{H} + \frac{1}{2}\comm{\rmi\mathcal{S}}{\comm{\rmi\mathcal{S}}{H}} + \ldots\notag\\
&= H_0 + \sum_{\angled{ij}}{T_{ij}} + \comm{\rmi\mathcal{S}}{H_0 + \sum_{\angled{ij}}{T_{ij}}}\notag\\
&\phantom{{}=} + \frac{1}{2}\comm{\rmi\mathcal{S}}{\comm{\rmi\mathcal{S}}{H_0 + \sum_{\angled{ij}}{T_{ij}}}} + \ldots
\end{align}
If we can expand the canonical transformation operator $\rmi\mathcal{S}$ in powers of $J/U$ as $\rmi\mathcal{S} = \rmi\mathcal{S}^1 + \rmi\mathcal{S}^2 + \ldots$, where $\rmi\mathcal{S}^m \sim (J / U)^m$, then, up to order $J^2/U$, the Hamiltonian is given by
\begin{align}
H^{\ast} = &H_0 + \sum_{\angled{ij}}{T_{ij}} + \comm{\rmi\mathcal{S}^1}{H_0} + \comm{\rmi\mathcal{S}^2}{H_0}\notag\\
& + \comm{\rmi\mathcal{S}^1}{\sum_{\angled{ij}}{T_{ij}}} + \frac{1}{2}\comm{\rmi\mathcal{S}^1}{\comm{\rmi\mathcal{S}^1}{H_0}}.\label{eq:Hcan1}
\end{align}

The form of the canonical transformation was chosen such that in the absence of disorder the high-energy terms in $H^{\ast}$ would cancel. Therefore $\rmi\mathcal{S}^1$ was chosen such that $\comm{\rmi\mathcal{S}^1}{H_0}$ cancels high-energy terms in $\sum_{\angled{ij}}{T_{ij}}$, leaving only hopping terms of the form $T_{ij}^{n,n+1}$ which represent low-energy hopping processes. And $\rmi\mathcal{S}^2$ was chosen such that $\comm{\rmi\mathcal{S}^2}{H_0} + \frac{1}{2}\comm{\rmi\mathcal{S}^1}{\comm{\rmi\mathcal{S}^1}{H_0}}$ cancels high-energy terms in $\comm{\rmi\mathcal{S}^1}{\sum_{\angled{ij}}{T_{ij}}}$. Using the identity $\comm{H_0}{T_{ij}^{nm}} = \varepsilon_{ij}^{nm}T_{ij}^{nm}$, we can obtain
\begin{equation}
\rmi\mathcal{S}^1 = \sum_{\angled{ij}}\sum_{n\ne m-1}\frac{T_{ij}^{nm}}{\varepsilon_0^{nm}}\label{eq:S1}
\end{equation}
and
\begin{align}
\rmi\mathcal{S}^2 &= \sum_{\angled{ij}\angled{kl}}{\sum_{\substack{n\ne m-1\\p}}{\frac{1}{\left(\varepsilon_0^{nm}\right)^2}\comm{T_{ij}^{nm}}{T_{kl}^{p,p+1}}}}\notag\\
&\qquad + \frac{1}{2}\sum_{\angled{ij}\angled{kl}}{\sum_{\substack{n\ne m-1\\p\ne q-1\\n-m\ne p-q}}{\frac{\comm{T_{ij}^{nm}}{T_{kl}^{q-1,p+1}}}{\varepsilon_0^{nm}\left(\varepsilon_0^{nm} + \varepsilon_0^{q-1,p+1}\right)}}}.\label{eq:S2}
\end{align}

\section{\label{app:gsenergy}The Ground State Energy}

The ground state energy, $\mathcal{E}$, is given by a sum of various terms given below. We have $\mathcal{E} = \calE{1}{0} + \sum_{i=1}^{2}\calE{i}{1} + \sum_{i=1}^{2}\calE{i}{2} + \sum_{i=1}^{4}\calE{i}{3} + \sum_{i=1}^{8}\calE{i}{4} + \sum_{i=1}^{8}\calE{i}{5}$. These terms are obtained by taking the expectation values of the terms of the canonically-transformed Hamiltonian in the Gutzwiller state, and considering the various possibilities for the site indices ($i,j,k,l$) that give rise to different hopping processes. The terms are as follows.

The interaction energy is given by
\begin{align}
\calE{1}{0} = \sum_{ni}\left(\frac{1}{2}U_i n(n - 1) - \mu n\right)\abs{\f{i,n}}^2.
\end{align}
The nearest-neighbour hopping is given by
\begin{align}
\calE{1}{1} = -\sum_{n\angled{ij}}J_{ij}(n+1)\fcc{i,n+1}\fcc{j,n}\f{i,n}\f{j,n+1}
\end{align}
and
\begin{align}
\calE{2}{1} = &\sum_{\angled{ij}}\sum_{n\ne m-1}\frac{J_{ij}}{U_0}\frac{\sqrt{m(n+1)}\delta\varepsilon_{ij}^{nm}}{n-m+1}\notag\\
&\times\fcc{i,n+1}\fcc{j,m-1}\f{i,n}\f{j,m}.
\end{align}
The density-density interaction energy is given by
\begin{align}
\calE{1}{2} = &\frac{1}{2}\sum_{\angled{ij}}\sum_{n\ne m-1}\frac{J_{ij}J_{ji}}{U_0}\frac{m(n+1)}{n-m+1}\notag\\
&\times\left(\abs{\f{i,n+1}}^2\abs{\f{j,m-1}}^2 - \abs{\f{i,n}}^2\abs{\f{j,m}}^2\right)
\end{align}
and
\begin{align}
\calE{2}{2} = &\frac{1}{2}\sum_{\angled{ij}}\sum_{n\ne m-1}\frac{J_{ij}J_{ji}}{U_0^2}\frac{g_{nm}g_{m-1,n+1}\delta\varepsilon_{ij}^{nm}}{(n-m+1)^2}\notag\\
&\times\left(\abs{\f{i,n+1}}^2\abs{\f{j,m-1}}^2 - \abs{\f{i,n}}^2\abs{\f{j,m}}^2\right).
\end{align}
The hopping terms where two bosons hop to the nearest neighbour are given by
\begin{align}
\calE{1}{3} = &\frac{1}{2}\sum_{n\angled{ij}}\frac{J_{ij}^2}{U_0}(n+1)\big(n\fcc{i,n+1}\fcc{j,n-1}\f{i,n-1}\f{j,n+1}\notag\\
& + (n+2)\fcc{i,n+2}\fcc{j,n}\f{i,n}\f{j,n+2}\big),
\end{align}
\begin{align}
\calE{2}{3} = &-\frac{1}{2}\sum_{\angled{ij}}\frac{J_{ij}^2}{U_0^2}\Biggl(\sum_{\substack{n\ne m-1\\n \ne m+1}}\frac{g_{nm}g_{n-1,m+1}\delta\varepsilon_{ij}^{n-1,m+1}}{(n-m+1)(n-m-1)}\notag\\
&\times\fcc{i,n+1}\fcc{j,m-1}\f{i,n-1}\f{j,m+1}\notag\\
& - \sum_{\substack{n\ne m-1\\n \ne m-3}}\frac{g_{nm}g_{n+1,m-1}\delta\varepsilon_{ij}^{n+1,m-1}}{(n-m+1)(n-m+3)}\notag\\
&\times\fcc{i,n+2}\fcc{j,m-2}\f{i,n}\f{j,m}\Biggr),
\end{align}
\begin{align}
\calE{3}{3} = &-\frac{1}{4}\sum_{n\angled{ij}}\frac{J_{ij}^2}{U_0^2}\sqrt{n+1}\Biggl(n\sqrt{n-1}\left(\delta\varepsilon_{ij}^{n,n-1} + \delta\varepsilon_{ij}^{n-1,n}\right)\notag\\
&\times\fcc{i,n+1}\fcc{j,n-2}\f{i,n-1}\f{j,n} \notag\\
&- (n+2)\sqrt{n+3}\left(\delta\varepsilon_{ij}^{n,n+3} + \delta\varepsilon_{ij}^{n+1,n+2}\right)\notag\\
&\times\fcc{i,n+2}\fcc{j,n+1}\f{i,n}\f{j,n+3}\Biggr)
\end{align}
and
\begin{align}
\calE{4}{3} = &-\frac{1}{4}\sum_{\angled{ij}}\frac{J_{ij}^2}{U_0^2}\notag\\
&\times\Biggl(\sum_{\substack{m\ne n\ne m-1\\n\ne m+1}}\frac{g_{nm}g_{n-1,m+1}\left(\delta\varepsilon_{ij}^{nm} + \delta\varepsilon_{ij}^{n-1,m+1}\right)}{(n-m)(n-m+1)}\notag\\
&\times\fcc{i,n+1}\fcc{j,m-1}\f{i,n-1}\f{j,m+1}\notag\\
& - \sum_{\substack{m-2\ne n\ne m-1\\n\ne m-3}}\frac{g_{nm}g_{n+1,m-1}\left(\delta\varepsilon_{ij}^{nm} + \delta\varepsilon_{ij}^{n+1,m-1}\right)}{(n-m+1)(n-m+2)}\notag\\
&\times\fcc{i,n+2}\fcc{j,m-2}\f{i,n}\f{j,m}\Biggr).
\end{align}
The hopping terms where two bosons from two different neighbouring sites hop to and from a site are given by
\begin{align}
\calE{1}{4} = &\frac{1}{2}\sum_{\angled{ij}\angled{ik}}\sum_{n\ne m-1}\frac{J_{ij}J_{ik}}{U_0}\frac{g_{nm}}{n-m+1}\notag\\
&\times\Bigl(g_{n-1,2n-m+1}\fcc{i,n+1}\fcc{j,m-1}\fcc{k,2n-m}\f{i,n-1}\notag\\
&\times\f{j,m}\f{k,2n-m+1} - g_{n+1,2n-m+3}\fcc{i,n+2}\fcc{j,m-1}\notag\\
&\times\fcc{k,2n-m+2}\f{i,n}\f{j,m}\f{k,2n-m+3}\Bigr),
\end{align}
\begin{align}
\calE{2}{4} = &\frac{1}{2}\sum_{\angled{ij}\angled{kj}}\sum_{n\ne m-1}\frac{J_{ij}J_{kj}}{U_0}\frac{\sqrt{m(n+1)}}{n-m+1}\notag\\
&\times\Bigl(g_{2m-n-1,m+1}\fcc{i,n+1}\fcc{j,m-1}\fcc{k,2m-n}\f{i,n}\notag\\
&\times\f{j,m+1}\f{k,2m-n-1} - g_{2m-n-3,m-1}\fcc{i,n+1}\fcc{j,m-2}\notag\\
&\times\fcc{k,2m-n-2}\f{i,n}\f{j,m}\f{k,2m-n-3}\Bigr),
\end{align}
\begin{align}
\calE{3}{4} = &-\frac{1}{2}\sum_{\angled{ij}\angled{ik}}\sum_{n\ne m-1}\frac{J_{ij}J_{ik}}{U_0^2}\frac{g_{nm}}{n-m+1}\notag\\
&\times\Biggl(\sum_{n\ne p+1}\frac{g_{n-1,p+1}\delta\varepsilon_{ik}^{n-1,p+1}}{n-p-1}\notag\\
&\times\fcc{i,n+1}\fcc{j,m-1}\fcc{k,p}\f{i,n-1}\f{j,m}\f{k,p+1}\notag\\
& - \sum_{n\ne p-1}\frac{g_{n+1,p+1}\delta\varepsilon_{ik}^{n+1,p+1}}{n-p+1}\notag\\
&\times\fcc{i,n+2}\fcc{j,m-1}\fcc{k,p}\f{i,n}\f{j,m}\f{k,p+1}\Biggr),
\end{align}
\begin{align}
\calE{4}{4} = &-\frac{1}{2}\sum_{\angled{ij}\angled{kj}}\sum_{n\ne m-1}\frac{J_{ij}J_{kj}}{U_0^2}\frac{g_{nm}}{n-m+1}\notag\\
&\times\Biggl(\sum_{q\ne m+1}\frac{g_{q-1,m+1}\delta\varepsilon_{kj}^{q-1,m+1}}{q-m-1}\notag\\
&\times\fcc{i,n+1}\fcc{j,m-1}\fcc{k,q}\f{i,n}\f{j,m+1}\f{k,q-1}\notag\\
&- \sum_{q\ne m-1}\frac{g_{q-1,m-1}\delta\varepsilon_{kj}^{q-1,m-1}}{q-m+1}\notag\\
&\times\fcc{i,n+1}\fcc{j,m-2}\fcc{k,q}\f{i,n}\f{j,m}\f{k,q-1}\Biggr),
\end{align}
\begin{align}
\calE{5}{4} = &-\sum_{\angled{ij}\angled{ik}}\sum_{n\ne m-1}\frac{J_{ij}J_{ik}}{U_0^2}\frac{g_{nm}}{(n-m+1)^2}\notag\\
&\times\Biggl(n\left(\delta\varepsilon_{ij}^{nm} + \delta\varepsilon_{ik}^{n-1,n}\right)\notag\\
&\times\fcc{i,n+1}\fcc{j,m-1}\fcc{k,n-1}\f{i,n-1}\f{j,m}\f{k,n}\notag\\
& - (n+2)\left(\delta\varepsilon_{ij}^{nm} + \delta\varepsilon_{ik}^{n+1,n+2}\right)\notag\\
&\times\fcc{i,n+2}\fcc{j,m-1}\fcc{k,n+1}\f{i,n}\f{j,m}\f{k,n+2}\Biggr),
\end{align}
\begin{align}
\calE{6}{4} = &-\sum_{\angled{ij}\angled{kj}}\sum_{n\ne m-1}\frac{J_{ij}J_{kj}}{U_0^2}\frac{g_{nm}}{(n-m+1)^2}\notag\\
&\times\Biggl((m+1)\left(\delta\varepsilon_{ij}^{nm} + \delta\varepsilon_{kj}^{m,m+1}\right)\notag\\
&\times\fcc{i,n+1}\fcc{j,m-1}\fcc{k,m+1}\f{i,n}\f{j,m+1}\f{k,m}\notag\\
& - (m-1)\left(\delta\varepsilon_{ij}^{nm} + \delta\varepsilon_{kj}^{m-2,m-1}\right)\notag\\
&\times\fcc{i,n+1}\fcc{j,m-2}\fcc{k,m-1}\f{i,n}\f{j,m}\f{k,m-2}\Biggr),
\end{align}
\begin{align}
\calE{7}{4} = &-\frac{1}{2}\sum_{\angled{ij}\angled{ik}}\sum_{n\ne m-1}\frac{g_{nm}}{n-m+1}\frac{J_{ij}J_{ik}}{U_0^2}\notag\\
&\times\Biggl(\sum_{\substack{p\ne n-1\\p\ne 2n-m}}\frac{g_{n-1,p+1}}{2n-m-p}\left(\delta\varepsilon_{ij}^{nm} + \delta\varepsilon_{ik}^{n-1,p+1}\right)\notag\\
&\times\fcc{i,n+1}\fcc{j,m-1}\fcc{k,p}\f{i,n-1}\f{j,m}\f{k,p+1}\notag\\
& - \sum_{\substack{p\ne n+1\\p\ne 2n-m+2}}\frac{g_{n+1,p+1}}{2n-m-p+2}\left(\delta\varepsilon_{ij}^{nm} + \delta\varepsilon_{ik}^{n+1,p+1}\right)\notag\\
&\times\fcc{i,n+2}\fcc{j,m-1}\fcc{k,p}\f{i,n}\f{j,m}\f{k,p+1}\Biggr)
\end{align}
and
\begin{align}
\calE{8}{4} = &-\frac{1}{2}\sum_{\angled{ij}\angled{kj}}\sum_{n\ne m-1}\frac{g_{nm}}{n-m+1}\frac{J_{ij}J_{kj}}{U_0^2}\notag\\
&\times\Biggl(\sum_{\substack{q\ne m+1\\q\ne 2m-n}}\frac{g_{q-1,m+1}}{n-2m+q}\left(\delta\varepsilon_{ij}^{nm} + \delta\varepsilon_{kj}^{q-1,m+1}\right)\notag\\
&\times\fcc{i,n+1}\fcc{j,m-1}\fcc{k,q}\f{i,n}\f{j,m+1}\f{k,q-1}\notag\\
& - \sum_{\substack{q\ne m-1\\q\ne 2m-n-2}}\frac{g_{q-1,m-1}}{n-2m+q+2}\left(\delta\varepsilon_{ij}^{nm} + \delta\varepsilon_{kj}^{q-1,m-1}\right)\notag\\
&\times\fcc{i,n+1}\fcc{j,m-2}\fcc{k,q}\f{i,n}\f{j,m}\f{k,q-1}\Biggr).
\end{align}
Finally, the next-nearest-neighbour hopping terms are given by
\begin{align}
\calE{1}{5} = &\frac{1}{2}\sum_{\angled{ij}\angled{jk}}\sum_{n\ne m-1}\frac{J_{ij}J_{jk}}{U_0}\frac{m(n+1)}{n-m+1}\notag\\
&\times\fcc{i,n+1}\fcc{k,n}\f{i,n}\f{k,n+1}\left(\abs{\f{j,m-1}}^2 - \abs{\f{j,m}}^2\right),
\end{align}
\begin{align}
\calE{2}{5} = &\frac{1}{2}\sum_{\angled{ij}\angled{ki}}\sum_{n\ne m-1}\frac{J_{ij}J_{ki}}{U_0}\frac{m(n+1)}{n-m+1}\notag\\
&\times\fcc{j,m-1}\fcc{k,m}\f{j,m}\f{k,m-1}\left(\abs{\f{i,n+1}}^2 - \abs{\f{i,n}}^2\right),
\end{align}
\begin{align}
\calE{3}{5} = &\frac{1}{2}\sum_{\angled{ij}\angled{jk}}\sum_{\substack{n\ne m-1\\p\ne m-1}}\frac{J_{ij}J_{jk}}{U_0^2}\frac{g_{nm}g_{m-1,p+1}\delta\varepsilon_{jk}^{m-1,p+1}}{(n-m+1)(p-m+1)}\notag\\
&\times\fcc{i,n+1}\fcc{k,p}\f{i,n}\f{k,p+1}\left(\abs{\f{j,m-1}}^2 - \abs{\f{j,m}}^2\right),
\end{align}
\begin{align}
\calE{4}{5} = &\frac{1}{2}\sum_{\angled{ij}\angled{ki}}\sum_{\substack{n\ne m-1\\n\ne q-1}}\frac{J_{ij}J_{ki}}{U_0^2}\frac{g_{nm}g_{q-1,n+1}\delta\varepsilon_{ki}^{q-1,n+1}}{(n-m+1)(n-q+1)}\notag\\
&\times\fcc{j,m-1}\fcc{k,q}\f{j,m}\f{k,q-1}\left(\abs{\f{i,n+1}}^2 - \abs{\f{i,n}}^2\right),
\end{align}
\begin{align}
\calE{5}{5} = &-\sum_{\angled{ij}\angled{jk}}\sum_{n\ne m-1}\frac{J_{ij}J_{jk}}{U_0^2}\frac{mg_{nm}\left(\delta\varepsilon_{ij}^{nm} + \delta\varepsilon_{jk}^{m-1,m}\right)}{(n-m+1)^2}\notag\\
&\times\fcc{i,n+1}\fcc{k,m-1}\f{i,n}\f{k,m}\left(\abs{\f{j,m-1}}^2 - \abs{\f{j,m}}^2\right),
\end{align}
\begin{align}
\calE{6}{5} = &-\sum_{\angled{ij}\angled{ki}}\sum_{n\ne m-1}\frac{J_{ij}J_{ki}}{U_0^2}\notag\\
&\times\frac{(n+1)g_{nm}\left(\delta\varepsilon_{ij}^{nm} + \delta\varepsilon_{ki}^{n,n+1}\right)}{(n-m+1)^2}\notag\\
&\times\fcc{j,m-1}\fcc{k,n+1}\f{j,m}\f{k,n}\left(\abs{\f{i,n+1}}^2 - \abs{\f{i,n}}^2\right),
\end{align}
\begin{align}
\calE{7}{5} = &-\frac{1}{2}\sum_{\angled{ij}\angled{jk}}\sum_{\substack{p\ne n\ne m-1\\p\ne m-1}}\frac{J_{ij}J_{jk}}{U_0^2}\notag\\
&\times\frac{g_{nm}g_{m-1,p+1}\left(\delta\varepsilon_{ij}^{nm} + \delta\varepsilon_{jk}^{m-1,p+1}\right)}{(n-m+1)(n-p)}\notag\\
&\times\fcc{i,n+1}\fcc{k,p}\f{i,n}\f{k,p+1}\left(\abs{\f{j,m-1}}^2 - \abs{\f{j,m}}^2\right)
\end{align}
and
\begin{align}
\calE{8}{5} = &-\frac{1}{2}\sum_{\angled{ij}\angled{ki}}\sum_{\substack{n\ne m-1\\n\ne q-1\\q\ne m}}\frac{J_{ij}J_{ki}}{U_0^2}\notag\\
&\times\frac{g_{nm}g_{q-1,n+1}\left(\delta\varepsilon_{ij}^{nm} + \delta\varepsilon_{ki}^{q-1,n+1}\right)}{(n-m+1)(q-m)}\notag\\
&\times\fcc{j,m-1}\fcc{k,q}\f{j,m}\f{k,q-1}\left(\abs{\f{i,n+1}}^2 - \abs{\f{i,n}}^2\right).
\end{align}

\bibliographystyle{epj}
\bibliography{paper}

\end{document}